# A Sophisticated Framework for the Accurate Detection of Phishing Websites

Asif Newaz [1]*, Farhan Shahriyar Haq [2]

[1,2] Department of Electrical and Electronic Engineering, Islamic University of Technology, Gazipur, Bangladesh;

**Email:** eee.asifnewaz@iut-dhaka.edu [1], farhanshahriyar@iut-dhaka.edu [2]

**\* Corresponding Author:**

**Address:** Department of Electrical and Electronic Engineering, Islamic University of Technology, Gazipur, Bangladesh - 1704**.**

**Email:** eee.asifnewaz@iut-dhaka.edu

**Contact:** +8801880841119

**Abstract**

Phishing is an increasingly sophisticated form of cyberattack that is inflicting huge financial damage to corporations throughout the globe while also jeopardizing individuals' privacy. Attackers are constantly devising new methods of launching such assaults and detecting them has become a daunting task. Many different techniques have been suggested, each with its own pros and cons. While machine learning-based techniques have been most successful in identifying such attacks, they continue to fall short in terms of performance and generalizability. This paper proposes a comprehensive methodology for detecting phishing websites. The goal is to design a system that is capable of accurately distinguishing phishing websites from legitimate ones and provides generalized performance over a broad variety of datasets. A combination of feature selection, greedy algorithm, cross-validation, and deep learning methods have been utilized to construct a sophisticated stacking ensemble classifier. Extensive experimentation on four different phishing datasets was conducted to evaluate the performance of the proposed technique. The proposed algorithm outperformed the other existing phishing detection models obtaining accuracy of 97.49%, 98.23%, 97.48%, and 98.20% on dataset-1 (UCI Phishing Websites Dataset), dataset-2 (Phishing Dataset for Machine Learning: Feature Evaluation), dataset-3 (Phishing Websites Dataset), and dataset-4 (Web page phishing detection), respectively. The high accuracy values obtained across all datasets imply the models' generalizability and effectiveness in the accurate identification of phishing websites.

# 1. Introduction

With the exponential growth of online activities, especially in the social, professional, and financial sectors, cyber-attacks have become a serious worry for the security of people and businesses in recent years. Phishing is one of the most hazardous and prevalent forms of cyber-attacks. It involves the use of fraudulent e-mails and websites to trick unwary users into supplying sensitive information like personal identity data, social security numbers, and financial credentials. The most serious consequence of phishing attacks is the misuse of information through the invasion and illegal exploitation of confidential user data, which can result in financial or other losses for victims. Phishing has existed for more than three decades, and each year, a substantial number of people are swindled, resulting in financial damages [1]. The number of phishing attacks has risen tremendously in recent years. In July 2021, the Anti-Phishing Working Group (APWC) reported 260,642 phishing assaults in their "Phishing Activity Trends Report - Q3 2021," indicating that the number of attacks had more than doubled from early 2020 [2].

Phishing attacks can be carried out using a variety of methods, including websites, emails, malware, or using a combination of these. For example, a phishing website can be created to imitate a legitimate website in order to defraud that organization or its users. The attacker then shares these fake website URLs via spam emails, social media, or other communication mediums to many internet users. Upon clicking those links, users will get redirected to the phishing website and their sensitive information will be compromised. Different approaches have been adopted to detect phishing websites and prevent these kinds of attacks [3-5]. These approaches can be classified into the following categories:

- **List-based Approach:** The majority of browsers verify phishing or legitimate webpages using permitted or disapproved resource lists (e.g., URLs, digital certificates, domains, etc.). The list of sites containing approved resources is referred to as a whitelist [6], whereas the list of disapproved resources constitutes a blacklist [7]. This approach is predicated on a central database of previously identified phishing URLs and websites. This limits its ability to identify newly launched phishing websites.

- **Visual Similarity-based Approach**: These anti-phishing approaches compare the image of a suspicious website to a legitimate image database [8,9]. Based on the similarity between the website and the database, a similarity ratio is generated, which is then used to declare whether the website is legitimate or not. When the similarity ratio exceeds a particular level, the website is labeled as phishing; otherwise, it is recognized as normal.

- **Content-based Approach:** These techniques detect phishing attacks by analyzing features retrieved from the phishing website. These features are extracted from various sources like URL, website traffic, page source code, search engine, HTML and JavaScript-based Features, DNS, etc. Based on the features extracted, machine learning classifiers can be modeled to distinguish between fraudulent and legitimate websites.

Among the three, the content-based approach is far more promising and is capable of detecting new phishing websites. This approach has two particular research focus. One is to identify the features or indicators that can be utilized to distinguish legitimate websites from fraudulent ones. The other is to develop a classification algorithm that can accurately classify the websites. Feature extraction is a crucial step in the content-based approach. In order to train the classifiers, a large number of both fraudulent and legitimate website instances is required. These sample websites are then analyzed to extract different features that might aid in determining their nature. PhishTank is one example of a phishing website database. It is a community-driven website maintained by OpenDNS that allows anybody to submit, verify, track, and share phishing data [10]. Data collected from such a database is then utilized to extract necessary indicators. Data scientists and researchers derived many different features from those data. Compilation of such work can be found in several publicly accessible datasets on which we conducted our experiment.

In this study, we focus on developing a phishing website detection system that is highly accurate and robust. In that regard, we propose a sophisticated classification model based on stacked generalization. In general, given multiple machine learning models, the one providing the highest accuracy is chosen for the classification task. These base models are often termed weak learners. Weak learners are usually more susceptible to noise and prone to overfitting. To improve the performance of such classifiers, an ensemble of them can be formed which are more robust. Bagging or boosting are two popular ensemble approaches. They use a group of weak learners (generally the same learner) to create a stronger, aggregated one. These are homogeneous classification algorithms. Stacked generalization is another type of ensemble approach that uses a meta-learning algorithm to learn how to best combine the predictions from multiple machine learning models often referred to as base learners. The predictions from the base learners get stacked and used as features by the meta-learner to make the final prediction. This is a heterogeneous classification algorithm. The base models are trained on the input data, while the meta-learner is trained on the predictions made by the base learners. This way, the stacking model harnesses the capabilities of a variety of different machine learning algorithms and is capable of outperforming any single one of them.

However, merely compiling several models together for stacking does not guarantee better performance. Each machine learning algorithm learns from the data in its own way. The phishing datasets consist of many different features. However, not all the features are relevant or related to the target concept. Building machine learning models using irrelevant features gives rise to several problems such as overfitting, poor generalization, and increased computational complexity [11]. The model tends to perform poorly on new data if it is trained on irrelevant features since the information used to develop the model does not represent the class properly. Building a stacking classifier using such biased models would certainly degrade the performance of the meta-model. In order to prevent that, the base models first need to be trained on the most representative subset of features. We utilized the Recursive Feature Elimination (RFE) technique to acquire such a feature set [12]. RFE is a wrapper approach, implying that the feature selection algorithm is wrapped around a particular classifier. As a result, the feature subset obtained for different classifiers would be different. The goal is to obtain a suitable feature set that optimizes the performance of the classifiers, thus reducing complexity and ensuring better generalization. Once the classifiers have been trained on suitable feature sets, they can be used as base learners for the meta-classifier.

One thing to be careful about while building a stacking classifier is data leakage. The predictions from the base classifiers should come from a subset of the training data that was not used to train the base classifiers. Otherwise, the stacking classifier would provide over-optimistic results and perform poorly on unseen data. A proper way to address this is to use the k-fold cross-validation method. A 10-fold cross-validation scheme was utilized in this study. The training dataset was split into 10 different folds. 9 folds were used to train the base classifiers, while the remaining fold was used to generate predictions upon which the meta-classifier was trained. This way, the leaking of information is avoided and a well-generalized stacking model is formed.

The next step is to select which classifiers will serve as the base classifiers. Using a low-performing classifier as the base model is likely to degrade the performance of the stacking ensemble model. A common strategy is to select the top N models based on their accuracy. However, that does not ensure the optimal performance for the ensemble. Moreover, there is no definite way to select the value of N. The best combination of the base classifiers that would optimize the performance can be found by searching through all possible combinations of the base classifiers. This exhaustive search approach can become computationally intensive with the increasing number of base classifiers. To keep the computational cost at a minimum while optimizing the performance of the stacked model, a heuristic search-based approach was employed. We start with the best-performing classifier, then add another with it from the pool of the remaining base classifiers that improves the performance of the stacked model when paired with the first one. This way, a subset of classifiers is obtained that would be used as the base classifiers for the stacking ensemble model.

Finally, a meta-model is used that would be trained on the predictions made by the base classifiers. We used a Multilayer Perceptron model (MLP) as the meta-model for our stacking classifier. MLP is a type of Artificial Neural Network (ANN) that can learn complex non-linear relationships in the data. The MLP model was tuned to obtain the best configuration for the final prediction.

This way, a robust stacking ensemble classifier architecture is constructed and utilized to distinguish phishing websites from legitimate ones. To ensure a thorough evaluation, the performance of our proposed approach is tested on four different phishing website datasets collected from the UCI repository [13] and Mendeley Data [14, 15, 16]. In all four cases, the proposed algorithm produced better results than other machine learning classifiers. The result obtained from the proposed approach is also found to be higher than what has been reported by other researchers.

## 2. Literature Review

Phishing is an increasingly sophisticated form of cyberattack causing tremendous loss to many global organizations. Hence, countermeasures for detecting such phishing activities are actively being developed. Due to the continual development and dynamic nature of phishing websites on the web, traditional blacklist-based phishing website detection algorithms cannot always accurately detect whether new websites are legitimate or fraudulent [17-22]. Consequently, many intelligent phishing detection techniques have been presented to address the issue of the constant emergence of new phishing websites and to distinguish newly built phishing websites from legal websites. These techniques are primarily based on Machine Learning (ML) and have been extensively employed as detection tools for a variety of cyber security threats. Ali et. al [23] employed a Genetic Algorithm (GA) based feature selection mechanism to remove redundant features from the data and then trained a feed-forward neural network to make predictions. They reported an accuracy of 91.13% and a sensitivity of 90.79% using their proposed hybrid technique (GA+DNN) on the UCI dataset [13]. In another work [24], they proposed a Particle Swarm Optimization (PSO)-based feature weighting approach to identify fraudulent websites. The authors tested their technique with different machine learning algorithms and achieved classification accuracies of 96.43% for Back-Propagation Neural Network (BPNN), 92.19% for Support Vector Machine (SVM), 91.03% for Naïve Bayes (NB), 96.28% for C4.5, 96.83% for Random Forest (RF), and 96.32% for K-Nearest Neighbor (KNN). Alqahtani et. al [25] proposed a promising data mining approach integrating classification and association rules to classify websites. Using their proposed approach, the authors were able to attain an accuracy of 95.20% on the UCI dataset [13]. Mehdi et. al [26] utilized nonlinear regression based on harmony search (HS) and obtained an accuracy of 94.13% on the training set and 92.80% on the test set on the UCI dataset [13]. Alsariera et. al [27] proposed four meta-learning algorithms developed using the extra-tree base classifier. Each of the proposed models had a detection accuracy of around 97.4% on the UCI dataset [13]. Mohammed et. al [28] proposed an optimized stacking ensemble classifier for phishing website detection. The authors tuned the parameters of the base classifiers using a genetic algorithm. Then chose the top three best-performing classifiers as base learners and an SVM classifier as a meta-learner. The proposed method was tested on three different phishing datasets achieving 97.16% accuracy on the UCI dataset [13], 98.58% accuracy on the Mendeley dataset [14], and 97.39% accuracy on another dataset collected from Mendeley [15]. Although the authors were able to obtain high accuracy using their proposed approach, optimizing the base learners using GA is computationally intensive. Each classifier has many different tunable parameters. While tuning them would certainly enhance the performance of the classifiers, the model loses its generalizability. The parameters need to be optimized for each dataset separately. As the parameters of the classifier are tuned for a particular set of data, the model tends to perform poorly on unseen data. So, the high accuracy obtained can be considered over-optimistic in that regard. Adeyemo et. al [29] proposed an ensemble-based Logistic model tree approach which is a combination of logistic regression and tree induction method into a single tree. They reported an accuracy of 97.18% on the UCI dataset [13].

In this study, we propose a comprehensive model for phishing website detection which offers generalized performance over a variety of datasets. The proposed algorithm outperformed the other existing phishing detection models obtaining accuracy of 97.49%, 98.23%, 97.48%, and 98.20% on dataset-1 (UCI Phishing Websites Dataset), dataset-2 (Phishing Dataset for Machine Learning: Feature Evaluation), dataset-3 (Phishing Websites Dataset), and dataset-4 (Web page phishing detection), respectively.

## 3. Materials and Methods

### 3.1 Dataset Description

To ensure the generalizability of the proposed system, a rigorous experiment was conducted on four different publicly available phishing datasets. These datasets are prepared and shared by researchers and data scientists for experimental purposes. One of them is collected from the UCI machine learning repository [13], while the others are collected from Mendeley data [14-16]. A summary of the datasets is provided in Table 1. A detailed description of the features of those datasets can be found on their source website. A brief description of the features of dataset-1 is outlined in Table 2.

Table 1: Summary of phishing datasets utilized in this study

| # | Dataset Name | Instances | Legitimate website instances | Phishing website instances | Feature |
|---|---|---|---|---|---|
| Dataset-1 | UCI Phishing Websites Dataset [13] | 11055 | 6157 | 4898 | 30 |
| Dataset-2 | Phishing Dataset for Machine Learning: Feature Evaluation [14] | 10000 | 5000 | 5000 | 48 |
| Dataset-3 | Phishing Websites Dataset [15] | 88647 | 58000 | 30647 | 111 |
| Dataset-4 | Web page phishing detection [16] | 11430 | 5715 | 5715 | 87 |

Table 2: Description of the features of Dataset-1

| Category | Serial | Feature name |
|---|---|---|
| Address Bar-based Features | 1 | Using the IP Address |
| | 2 | Long URL to Hide the Suspicious Part |
| | 3 | Using URL Shortening Services "TinyURL" |
| | 4 | URL's having "@" Symbol |
| | 5 | Redirecting using "//" |
| | 6 | Adding Prefix or Suffix Separated by (-) to the Domain |
| | 7 | Sub Domain and Multi-Sub Domains |
| | 8 | HTTPS (Hyper Text Transfer Protocol with Secure Sockets Layer) |
| | 9 | Domain Registration Length |
| | 10 | Favicon |
| | 11 | Using Non-Standard Port |
| | 12 | The Existence of "HTTPS" Token in the Domain Part of the URL |
| Abnormal Based Features | 13 | Request URL |
| | 14 | URL of Anchor |

|  | 15 | Links in <Meta>, <Script> and <Link> tags |
|---|---|---|
|  | 16 | Server Form Handler (SFH) |
|  | 17 | Submitting Information to Email |
|  | 18 | Abnormal URL |
| HTML and JavaScript-based Features | 19 | Website Forwarding |
|  | 20 | Status Bar Customization |
|  | 21 | Disabling Right Click |
|  | 22 | Using Pop-up Window |
|  | 23 | IFrame Redirection |
| Domain-based Features | 24 | Age of Domain |
|  | 25 | DNS Record |
|  | 26 | Website Traffic |
|  | 27 | PageRank |
|  | 28 | Google Index |
|  | 29 | Number of Links Pointing to Page |
|  | 30 | Statistical-Reports Based Feature |

## 3.2 Stack Generalization

Stack Generalization is an ensemble learning technique where the predictions from multiple classifiers (generally referred to as base classifier or level-1 classifier) are used as variables to train a meta classifier (level-2 classifier). The base classifiers are trained on the training set and the final predictions are made on the test set by the meta-classifier. Unlike homogeneous ensemble classifiers where the base classifiers belong to the same category (Bagged or Boosted trees), the stacking classifier is a heterogeneous ensemble classifier that can combine the predictions from a wide range of classifiers. This enables the classifier to learn more complex patterns and provide more accurate performance compared to other algorithms. The configuration of a standard stacking classifier is illustrated in Figure 1.

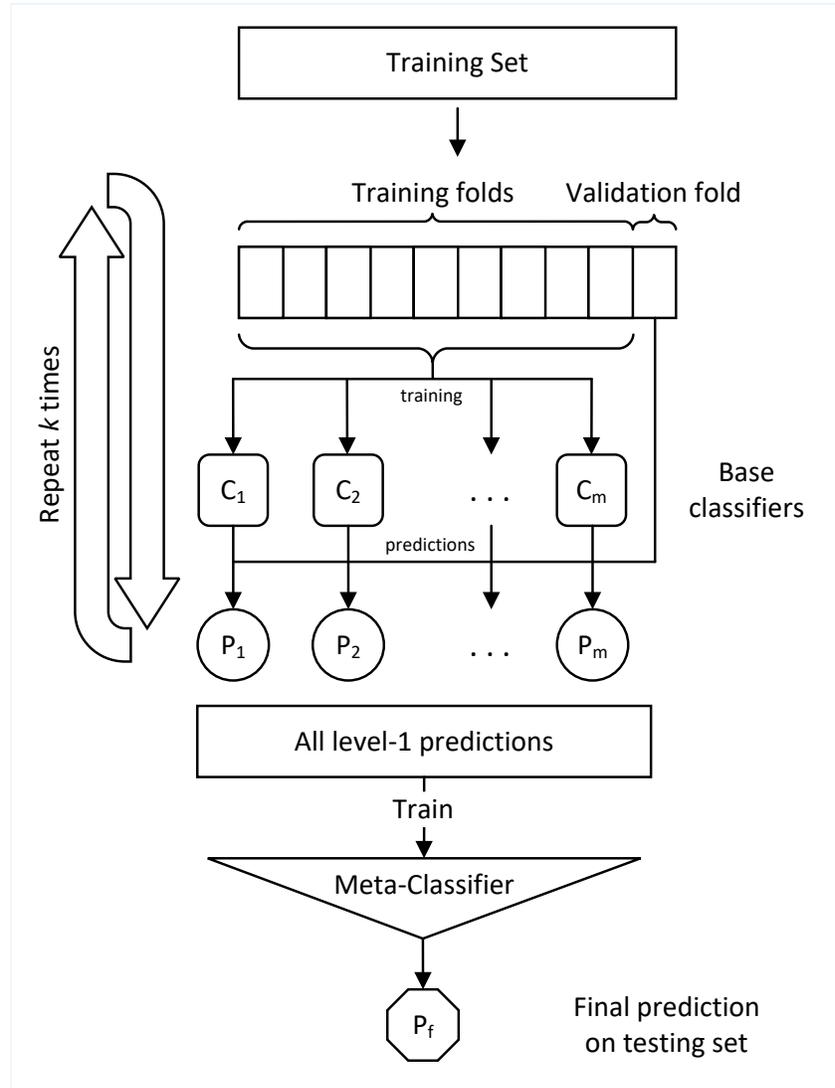

Figure 1: Configuration of the Stacking Ensemble Classifier

### 3.3 Proposed Methodology

The aim of this study is to develop a reliable detection system that can discern phishing websites from legitimate ones. It is crucial to identify phishing websites correctly. However, it is also important to minimize the number of false detections as it can cause dissatisfaction. In that regard, a sophisticated phishing detection framework based on the stacking ensemble classifier has been presented in this literature. The architecture of the proposed framework is illustrated in Figure 2. Rigorous experimentation has been performed on four different phishing datasets to ensure proper evaluation. Six different performance measures – accuracy, sensitivity, precision, g-mean Score, F1-score, and ROC-AUC were utilized to properly assess the performance of the proposed method.

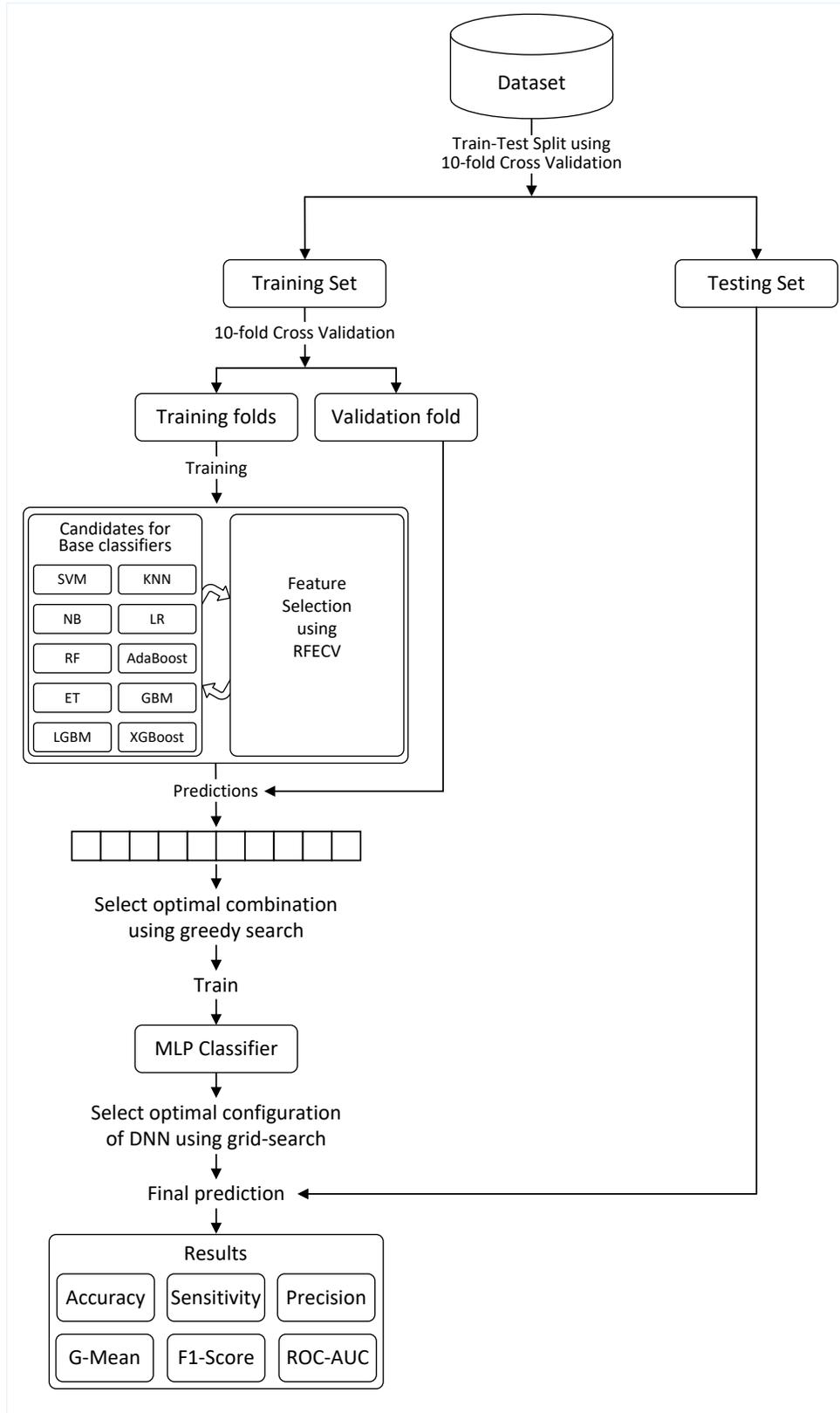

Figure 2: Outline of the proposed framework

### 3.3.1     Data Preparation

The datasets were first prepared before feeding them to the model. The categorical variables were converted into numeric variables. Then the entire dataset was divided into 2 parts – training (90%) and testing (10%) set using k-fold cross-validation. The base classifiers were trained on the training set and the final prediction from the stacking classifier was made on the test set. To ensure generalizability, this process was repeated 10 times with random splitting of data into train and test sets. The average results for 10 test sets were taken.

### 3.3.2     Construction of the ensemble classifier

The following steps were taken to construct a robust stacking ensemble classifier –

i. A k-fold cross-validation strategy was undertaken to split the training set into training and validation folds. The base classifiers were trained on the training folds, while the meta-classifier was trained on the predictions made by the base classifiers on the validation fold. This strategy is undertaken to avoid any data leakage that would produce over-optimistic results. The value of 'k' was taken as 10 in this study.

ii. Experimentation was performed on 10 classification algorithms – Support Vector Machine (SVM), K-Nearest Neighbors (KNN), Naïve Bayes (NB), Logistic Regression (LR), Random Forest (RF), AdaBoost, Extra Tree (ET), Gradient Boosting Machine (GBM), Light Gradient Boosting Machine (LGBM) and Extreme Gradient Boosting (XGBoost). They were considered a suitable candidate for the base classifier.

iii. Training the classifiers with the most representative subset of features can boost the classifier's performance while also reducing the complexity of the system. Therefore, before training, feature selection was performed using Recursive Feature Elimination with Cross-Validation (RFECV) technique [30]. RFECV is a popular wrapper algorithm and an extension of the RFE technique. Here, cross-validation is utilized to automatically tune the number of features to be selected. The core idea behind the approach is to recursively eliminate the features which are less important than the others. This is accomplished by training the classifier on the initial set of features and calculating the importance of each feature. Then the least important feature is pruned from the current set of features. In the case of the datasets where the number of features is large, three features were removed at once to accelerate the process. The process is recursively repeated on the pruned set until the model stops improving.

iv. Using the RFECV approach, the suitable feature set for each classifier is obtained. The features obtained for different classifiers may vary from one another. Using the obtained set of features, each classifier is then trained on the training folds, and prediction was made on the validation fold. The process is repeated for each unique group among the 10 folds.

v. In general, the predictions made by the classifiers are binary (True or False), which is in this case whether the website is legitimate or not. The decision is taken based on class probability (p). A limit is set, which is generally 0.5. If the probability of the output variable is higher than that ($p > 0.5$), the classifier would predict it as true. Otherwise, if $p <= 0.5$, the prediction would be false. The flaw in this design is that the output probability of 0.95 or 0.55, both are considered equally true which is not entirely accurate. However, for the stack generalization, the predictions made by the base classifiers are not the final prediction, instead, they are to be used by the meta-classifier to provide the ultimate decision. On that account, instead of using binary prediction values from the base classifiers, the class probabilities of the output variables were utilized. The meta-classifier was trained using these class probabilities. This enables the meta-classifier to provide a more concrete decision.

vi. The next important thing in the construction of the ensemble is to choose the set of classifiers that would be used as base classifiers for the stacking model. Using any classifier that does not contribute to improving the performance of the meta-classifier would only increase the complexity of the system. Furthermore, it could also play a negative role by degrading the performance of the ensemble. Therefore, it is crucial to select the base classifiers carefully. On that account, the greedy algorithm was employed as the selection mechanism.

vii. Among the ten candidate classifiers, the best performing one was first added to a pool of base classifiers that are to be used for building the stacked model. Then from the remaining candidate classifiers, each one was added to the already selected classifier, and the performance of the stacked model with the two as the base classifier was observed. The combination that improved the performance of the stacked model was considered the new pool of base classifiers. The process is repeated, and a classifier is added to the pool until the model stops improving. This way, a subset of classifiers is obtained that optimizes the model's performance.

viii. Using the predictions made by the chosen set of base classifiers as features, a meta-learner was trained. Multilayer Perceptron (MLP) model was used as the meta-learner. MLP uses a deep neural network (DNN) architecture to learn from the data. The performance of the MLP classifier can vary with different configurations (i.e., the number of layers and neurons) of the DNN. As there is no definite rule for the configuration setup, the grid-search approach was undertaken to procure a suitable structure for the MLP classifier.

After the meta-learner is trained, it is used to make predictions on the test set which was kept apart at the data preparation stage (section – 3.3.1). The entire process (i – viii) is repeated and the prediction is then made on a different fold. The average of the performance on 10 different testing folds is considered and reported in the 'Results and Discussion' section.

## 4. Results and Discussion

Initially, using a 10-fold cross-validation technique, the datasets were split into 10 different folds. 9 of the 10 folds were considered as training sets and the remaining fold was used as the test set. This was done for each unique set of train-test folds. The results reported here are the average of the performance on 10 unique test sets.

### 4.1 Performance of 10 classification algorithms on four phishing datasets

10 different classification algorithms are considered as the candidate for the base classifiers. The performance of those classifiers on the four phishing datasets is reported in Tables 3 to 6. As can be observed, the ensemble algorithms like ET or XGBoost classifiers produced better results compared to other algorithms like SVM or NB. The NB algorithm produced the lowest accuracy in all datasets. The extra Tree classifier produced the highest accuracy of 97.37% in dataset-1. In dataset-2, the XGBoost classifier provided a maximum accuracy of 98.15%. In dataset-3, the highest accuracy (97.16%) was obtained by the RF classifier. XGBoost classifier provided the maximum accuracy of 97.08% on dataset-4. Although the accuracies obtained are quite high, one thing to consider here is that the same classifier did not provide the best performance in all datasets. So, choosing a particular classifier might not be able to provide optimal performance in a practical scenario. Moreover, due to the large number of samples on the datasets, even an accuracy of 97% still implies that there are hundreds of examples that were inaccurately predicted which is undesirable. This necessitates the development of an algorithm capable of minimizing the number of wrong predictions, while also providing optimal performance in all cases. In that regard, a framework for phishing website prediction has been proposed in this study based on stack generalization.

Table 3: Performance (in percentage) of the classifiers on Dataset-1

| Classifiers | Accuracy | Sensitivity | Precision | G-mean | F1-score | ROC-AUC |
| --- | --- | --- | --- | --- | --- | --- |
| SVM | 94.74 | 96.76 | 93.98 | 94.45 | 95.35 | 94.48 |
| KNN | 93.59 | 93.62 | 94.82 | 93.59 | 94.21 | 93.59 |
| NB | 60.39 | 28.99 | 99.60 | 53.77 | 44.87 | 64.42 |
| LR | 92.53 | 94.40 | 92.38 | 92.27 | 93.37 | 92.30 |
| RF | 97.28 | 98.23 | 96.93 | 97.14 | 97.57 | 97.15 |
| AdaBoost | 93.25 | 95.34 | 92.76 | 92.95 | 94.03 | 92.98 |
| ET | 97.37 | 97.90 | 97.39 | 97.29 | 97.64 | 97.29 |
| GBM | 94.70 | 96.08 | 94.50 | 94.51 | 95.28 | 94.52 |
| LGBM | 96.88 | 97.68 | 96.76 | 96.77 | 97.21 | 96.77 |
| XGBoost | 97.29 | 98.20 | 96.98 | 97.16 | 97.58 | 97.17 |

Table 4: Performance (in percentage) of the classifiers on Dataset-2

| Classifiers | Accuracy | Sensitivity | Precision | G-mean | F1-score | ROC-AUC |
| --- | --- | --- | --- | --- | --- | --- |
| SVM | 83.68 | 90.58 | 81.08 | 82.76 | 85.07 | 83.68 |
| KNN | 83.35 | 86.04 | 82.35 | 83.06 | 83.88 | 83.35 |
| NB | 84.26 | 74.16 | 92.91 | 83.61 | 82.41 | 84.26 |
| LR | 93.66 | 94.42 | 93.06 | 93.64 | 93.72 | 93.66 |
| RF | 97.72 | 97.2 | 98.22 | 97.72 | 97.71 | 97.72 |
| AdaBoost | 96.76 | 97.20 | 96.37 | 96.76 | 96.77 | 96.76 |
| ET | 97.36 | 96.04 | 98.64 | 97.35 | 97.32 | 97.36 |
| GBM | 97.25 | 97.22 | 97.28 | 97.25 | 97.25 | 97.25 |
| LGBM | 98.02 | 97.72 | 98.32 | 98.02 | 98.02 | 98.02 |
| XGBoost | 98.15 | 97.84 | 98.46 | 98.14 | 98.14 | 98.15 |

Table 5: Performance (in percentage) of the classifiers on Dataset-3

| Classifiers | Accuracy | Sensitivity | Precision | G-mean | F1-score | ROC-AUC |
|---|---|---|---|---|---|---|
| SVM | 75.79 | 56.72 | 67.97 | 69.79 | 61.84 | 71.30 |
| KNN | 86.64 | 76.35 | 83.59 | 83.84 | 79.81 | 84.22 |
| NB | 84.28 | 62.80 | 88.38 | 77.50 | 73.43 | 79.22 |
| LR | 89.34 | 79.73 | 88.35 | 86.69 | 83.69 | 87.07 |
| RF | 97.16 | 96.13 | 95.69 | 96.92 | 95.91 | 96.92 |
| AdaBoost | 93.65 | 90.71 | 90.90 | 92.92 | 90.80 | 92.96 |
| ET | 97.09 | 95.93 | 95.67 | 96.81 | 95.80 | 96.82 |
| GBM | 95.35 | 93.46 | 93.11 | 94.89 | 93.28 | 94.90 |
| LGBM | 96.68 | 95.33 | 95.09 | 96.36 | 95.21 | 96.36 |
| XGBoost | 97.08 | 95.78 | 95.78 | 96.77 | 95.78 | 96.77 |

Table 6: Performance (in percentage) of the classifiers on Dataset-4

| Classifiers | Accuracy | Sensitivity | Precision | G-mean | F1-score | ROC-AUC |
|---|---|---|---|---|---|---|
| SVM | 95.91 | 95.55 | 96.23 | 95.90 | 95.89 | 95.90 |
| KNN | 93.44 | 90.88 | 95.79 | 93.41 | 93.27 | 93.45 |
| NB | 68.69 | 39.87 | 94.11 | 62.33 | 55.97 | 68.69 |
| LR | 94.48 | 94.06 | 94.85 | 94.47 | 94.45 | 94.47 |
| RF | 96.65 | 96.46 | 96.82 | 96.65 | 96.64 | 96.65 |
| AdaBoost | 94.63 | 94.31 | 94.93 | 94.63 | 94.62 | 94.64 |
| ET | 96.59 | 96.03 | 97.12 | 96.59 | 96.56 | 96.59 |
| GBM | 95.71 | 95.57 | 95.85 | 95.71 | 95.71 | 95.71 |
| LGBM | 96.84 | 96.81 | 96.87 | 96.84 | 96.84 | 96.84 |
| XGBoost | 97.08 | 97.21 | 96.94 | 97.08 | 97.08 | 97.08 |

## 4.2 Performance of the proposed approach on four phishing datasets

Stack generalization is an ensemble approach that harnesses the capabilities of a range of well-performing models. If the weak learners are properly combined, the stacking ensemble can outperform any single model and offer well-generalized performance. In that regard, a sophisticated architecture for constructing a robust stacked model is proposed in this literature. The construction process is described in detail in section - 3.3. The set of features that optimizes the performance of the classifiers was obtained using the RFECV technique. Among the candidate classifiers, the base classifiers were chosen using a greedy search algorithm. An MLP classifier was used to combine the predictions from the chosen classifiers. The performance of the proposed model on four phishing datasets is reported in Table 7.

As can be observed from Table 7, the proposed model produced the highest accuracy across all datasets. In terms of other metrics like the g-mean score, which indicates a balanced performance on both positive and negative cases, the stacked model outperformed the other classifiers as well. The classifier provided the maximum accuracy of 97.49% on dataset-1, 98.23% on dataset-2, 97.48% on dataset-3, and 98.20% on dataset-4. The precision (an important indicator of how accurately the framework is able to predict the positive cases) score achieved is also quite high, as high as 98.69% on dataset-2. Based on the results, it can be concluded that the suggested model is a well-generalized model capable of accurately identifying phishing websites.

Table 7: Performance (in percentage) of the proposed stacked model

| Dataset | Accuracy | Sensitivity | Precision | G-mean | F1-score | ROC-AUC |
|---|---|---|---|---|---|---|
| Dataset-1 | 97.49 | 98.43 | 97.11 | 97.35 | 97.76 | 97.35 |
| Dataset-2 | 98.23 | 97.71 | 98.69 | 98.20 | 98.20 | 98.21 |
| Dataset-3 | 97.48 | 98.43 | 97.11 | 97.34 | 97.75 | 97.36 |
| Dataset-4 | 98.20 | 94.06 | 94.85 | 94.47 | 94.45 | 94.47 |

The parameters of the base learners of the stacked model were not optimized in this work. Parameter optimization can improve the accuracy of the classifiers. However, it is an extremely time-consuming process that often leads to a loss of generalization. Therefore, parameter tuning was avoided, and the default parameters of the scikit-learn library were used to train the classifiers. As for the structure of the neural network, a few different combinations of layers were tested. A 5-layer deep neural network structure was found to be the optimal one for the meta-classifier. The same configuration was utilized for all datasets. The number of neurons on successive layers is – 6, 12, 32, 12, 6. RFECV algorithm was utilized to identify the most suitable subset of features for each classifier. The number of features on dataset-1 is only 30. However, other datasets like dataset-3 contain 111 features. RFECV algorithm was particularly effective in reducing the number of features in those cases. The base classifiers that were chosen from the ten candidate classifiers varied for different datasets. The selected base classifiers for dataset – 1 were: RF, ET, and XGBoost. The list of features selected for dataset-1 is outlined in Table 8.

Table 8: Features selected by RFECV technique on dataset-1

| Classifier Name | Selected Features |
| --- | --- |
| SVM | 1, 3, 4, 6, 7, 8, 11, 12, 13, 14, 15, 16, 17, 18, 19, 21, 22, 23, 25, 26, 27, 28, 29, 30 |
| KNN | 2, 3, 5, 6, 8, 9, 10, 12, 13, 14, 15, 18, 20, 23, 29 |
| NB | 2, 5, 7, 8, 9, 10, 11, 14, 15, 17, 19, 20, 22, 23, 29 |
| LR | 1, 3, 6, 7, 8, 10, 11, 12, 14, 15, 16, 17, 19, 25, 26, 28, 29,30 |
| RF | 1, 2, 3, 4, 5, 6, 7, 8, 9, 10, 11, 12, 13, 14, 15, 16, 17, 18, 19, 20, 21, 22, 23, 24, 25, 26, 27, 28, 29, 30 |
| AdaBoost | 1, 2, 3, 4, 6, 7, 8, 9, 10, 14, 15, 16, 19, 25, 26, 27, 28,29 |
| ET | 1, 2, 3, 4, 5, 6, 7, 8, 9, 10, 11, 12, 13, 14, 15, 16, 17, 18, 19, 20, 22, 23, 24, 25, 26, 27, 28, 29, 30 |
| GBM | 1, 2, 3, 5, 6, 7, 8, 9, 10, 12, 13, 14, 15, 16, 17, 18, 19, 20, 21, 22, 23, 24, 25, 26, 27, 28, 29 |
| LGBM | 1, 2, 3, 4, 5, 6, 7, 8, 9, 10, 12, 13, 14, 15, 16, 17, 18, 19, 20, 22, 23, 24, 25, 26, 27, 28, 29, 30 |
| XGBoost | 1, 2, 3, 6, 7, 8, 9, 10, 11, 12, 13, 14, 15, 16, 17, 18, 19, 20, 22, 23, 24, 25, 26, 27, 28, 29 |

The comparison in performance (in terms of accuracy) of the proposed approach with other classification algorithms is illustrated in Figure 3.

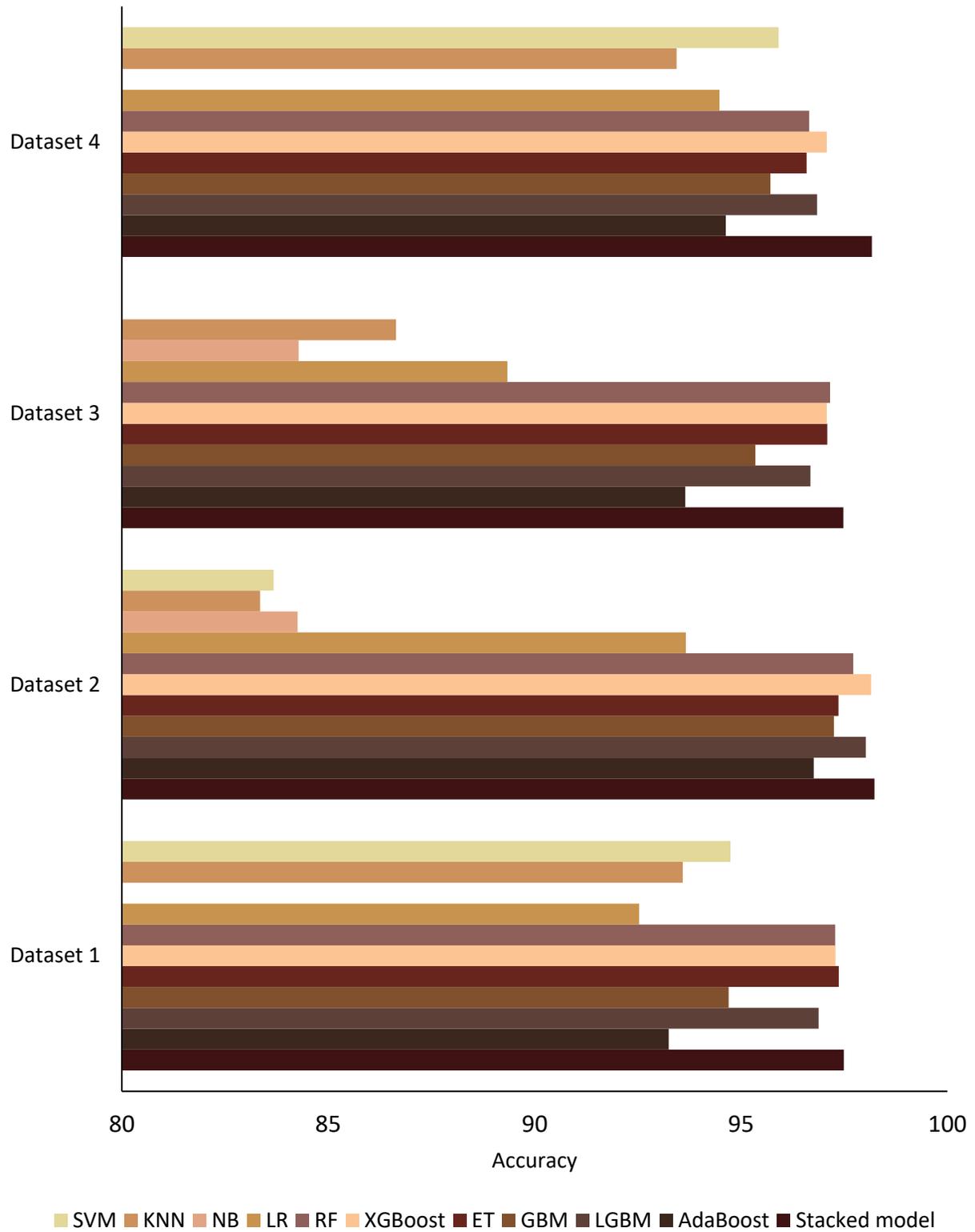

**Figure 3:** Performance comparison (in terms of accuracy) of the proposed approach with other machine learning algorithms

## 4.3 Performance comparison of the proposed approach with previous studies

Researchers have proposed many different techniques to detect phishing websites. A comparative analysis of the proposed approach with previous studies is provided in Table 9. As can be observed from the table, our proposed methodology has outperformed the other approaches in all cases. Dataset-4 has not been utilized in other studies. However, the proposed approach produced an accuracy of 98.20% which is considerably high.

**Table 9:** Performance comparison of the proposed approach with previous works

| Paper | Methodology | Dataset | Accuracy | Precision | Recall | F-Measure |
|---|---|---|---|---|---|---|
| Ali et. al [23] | Genetic algorithm (GA) + DNN | Dataset 1 | 91.13 | | 93.34 | |
| Ali et. al [24] | PSO – RF | Dataset 1 | 96.83 | 98.76 | 95.37 | |
| Alqahtani [25] | Association Rule | Dataset 1 | 95.20 | | | 0.951 |
| Adeyemo et. al [29] | ABLMT: AdaBoostLMT | Dataset 1 | 97.42 | 97.40 | 97.40 | 0.974 |
| | BGLMT: BaggingLMT | | 97.18 | 97.20 | 97.20 | 0.972 |
| Balogun et. al [31] | FT based meta-learning | Dataset 1 | 97.19 | | | 0.972 |
| | | Dataset 2 | 98.51 | | | 0.985 |
| Khan et. al [32] | ANN + PCA | Dataset 2 | 97.13 | | | |
| Lokesh et. al [33] | RF | Dataset 1 | 96.87 | | | |
| This Study | Robust Stacked Generalization | Dataset 1 | **97.49** | 97.11 | 98.43 | 97.76 |
| | | Dataset 2 | **98.23** | 98.69 | 97.71 | 98.20 |
| | | Dataset 3 | **97.48** | 97.11 | 98.43 | 97.75 |
| | | Dataset 4 | **98.20** | 94.85 | 94.06 | 94.45 |

## 5. Conclusion

This article presents a robust framework for phishing website detection. A 10-fold cross-validation strategy is undertaken to avoid data leakage while building the ensemble. Feature selection using recursive feature elimination is performed to obtain the most representable subset of features for each classification algorithm. 10 different classifiers are trained using the obtained feature set. The outputs from the algorithms are taken as probability values instead of binary to offer more flexibility while training the meta-classifier. A greedy algorithm-based selection mechanism is constructed to determine the combination of base learners that optimize the performance of the ensemble. A deep neural network architecture was used as the meta-classifier and the final predictions from the stacking ensemble classifier were validated on four phishing datasets. Thus, a sophisticated stacking ensemble classifier architecture is formed that produces well-generalized performance across all datasets. An accuracy of 97.49%, 98.23%, 97.48%, and 98.20% was obtained on the four datasets, respectively, which are comparatively higher than other classification algorithms. The proposed approach also outperformed the other existing models for phishing website detection. The high accuracy, as well as the versatile performance obtained on a diverse range of datasets, indicates the efficacy of the proposed approach and its potential for the accurate identification of phishing websites.

## 6. Data Availability Statement

All four of the datasets utilized in this study are publicly available in the UCI repository [13] and Mendeley data [14-16].

## 7. Conflicts of Interest Disclosure

We declare that there is no conflict of interest.

## 8. Sources of Funding

This research did not receive any specific grant from funding agencies in the public, commercial, or not-for-profit sectors.

## References

[1] Tang, L., & Mahmoud, Q. H. (2021). A survey of machine learning-based solutions for phishing website detection. Machine Learning and Knowledge Extraction, 3(3), 672-694. https://doi.org/10.3390/make3030034
[2] *APWG*. (n.d.). Apwg.Org. Retrieved April 2, 2022, from https://apwg.org/trendsreports/
[3] Afroz, S., & Greenstadt, R. (2011, September). Phishzoo: Detecting phishing websites by looking at them. In 2011 IEEE fifth international conference on semantic computing (pp. 368-375). IEEE. https://doi.org/10.1109/ICSC.2011.52
[4] Rao, R. S., & Pais, A. R. (2019). Detection of phishing websites using an efficient feature-based machine learning framework. Neural Computing and Applications, 31(8), 3851-3873. https://doi.org/10.1007/s00521-017-3305-0
[5] Jain, A. K., & Gupta, B. B. (2018). Towards detection of phishing websites on client-side using machine learning based approach. Telecommunication Systems, 68(4), 687-700. https://doi.org/10.1007/s11235-017-0414-0

[6] Cao, Y., Han, W., & Le, Y. (2008, October). Anti-phishing based on automated individual white-list. In Proceedings of the 4th ACM workshop on Digital identity management (pp. 51-60). https://doi.org/10.1145/1456424.1456434

[7] Prakash, P., Kumar, M., Kompella, R. R., & Gupta, M. (2010, March). Phishnet: predictive blacklisting to detect phishing attacks. In 2010 Proceedings IEEE INFOCOM (pp. 1-5). IEEE. https://doi.org/10.1109/INFCOM.2010.5462216

[8] Fu, A. Y., Wenyin, L., & Deng, X. (2006). Detecting phishing web pages with visual similarity assessment based on earth mover's distance (EMD). IEEE transactions on dependable and secure computing, 3(4), 301-311. https://doi.org/10.1109/TDSC.2006.50

[9] Chen, K. T., Chen, J. Y., Huang, C. R., & Chen, C. S. (2009). Fighting phishing with discriminative keypoint features. IEEE Internet Computing, 13(3), 56-63. https://doi.org/10.1109/MIC.2009.59

[10] *Phistank*. (n.d.). phistank.Org. Retrieved April 2, 2022, from https://phishtank.org/

[11] Bolón-Canedo, V., Sánchez-Maroño, N., & Alonso-Betanzos, A. (2015). Recent advances and emerging challenges of feature selection in the context of big data. Knowledge-based systems, 86, 33-45. https://doi.org/10.1016/j.knosys.2015.05.014

[12] *Sklearn.Feature_selection.RFE*. (n.d.). Scikit-Learn. Retrieved April 2, 2022, from https://scikit-learn.org/stable/modules/generated/sklearn.feature_selection.RFE.html

[13] *UCI machine learning repository: Phishing websites data set.* (n.d.). Uci.Edu. Retrieved April 2, 2022, from https://archive.ics.uci.edu/ml/datasets/phishing+websites

[14] Tan, C. L. (2018). *Phishing dataset for machine learning: Feature evaluation* [Data set]. Mendeley. https://doi.org/10.17632/h3cgnj8hft.1

[15] Vrbančič, G. (2020). *Phishing Websites Dataset* [Data set]. Mendeley. https://doi.org/10.17632/72ptz43s9v.1

[16] Hannousse, A. (2020). *Web page phishing detection* [Data set]. Mendeley. https://doi.org/10.17632/c2gw7fy2j4.1

[17] Mohammad, R. M., Thabtah, F., & McCluskey, L. (2014). Predicting phishing websites based on self-structuring neural network. *Neural Computing and Applications*, *25*(2), 443-458. https://doi.org/10.1007/s00521-013-1490-z

[18] He, M., Horng, S.-J., Fan, P., Khan, M. K., Run, R.-S., Lai, J.-L., Chen, R.-J., & Sutanto, A. (2011). An efficient phishing webpage detector. *Expert Systems with Applications*, *38*(10), 12018–12027. https://doi.org/10.1016/j.eswa.2011.01.046

[19] Mohammad, R. M., Thabtah, F., & McCluskey, L. (2015). Tutorial and critical analysis of phishing websites methods. *Computer Science Review*, *17*, 1–24. https://doi.org/10.1016/j.cosrev.2015.04.001

[20] Abdelhamid, N., Ayesh, A., & Thabtah, F. (2014). Phishing detection based associative classification data mining. *Expert Systems with Applications*, *41*(13), 5948-5959. https://doi.org/10.1016/j.eswa.2014.03.019

[21] Nguyen, H. H., & Nguyen, D. T. (2016). Machine learning based phishing web sites detection. In *AETA 2015: recent advances in electrical engineering and related sciences* (pp. 123-131). Springer, Cham. https://doi.org/10.1007/978-3-319-27247-4_11

[22] Ali, W. (2017). Phishing website detection based on supervised machine learning with wrapper features selection. *International Journal of Advanced Computer Science and Applications*, *8*(9), 72-78. https://doi.org/10.14569/IJACSA.2017.080910

[23] Ali, W., & Ahmed, A. A. (2019). Hybrid intelligent phishing website prediction using deep neural networks with genetic algorithm-based feature selection and weighting. *IET Information Security*, *13*(6), 659-669. https://doi.org/10.1049/iet-ifs.2019.0006

[24] Ali, W., & Malebary, S. (2020). Particle swarm optimization-based feature weighting for improving intelligent phishing website detection. *IEEE Access*, *8*, 116766-116780. https://doi.org/10.1109/ACCESS.2020.3003569

[25] Alqahtani, M. (2019, April). Phishing websites classification using association classification (PWCAC). In *2019 International conference on computer and information sciences (ICCIS)* (pp. 1-6). IEEE. https://doi.org/10.1109/ICCISci.2019.8716444

[26] Babagoli, M., Aghababa, M. P., & Solouk, V. (2019). Heuristic nonlinear regression strategy for detecting phishing websites. *Soft Computing*, *23*(12), 4315-4327. https://doi.org/10.1007/s00500-018-3084-2

[27] Alsariera, Y. A., Adeyemo, V. E., Balogun, A. O., & Alazzawi, A. K. (2020). Ai meta-learners and extra-trees algorithm for the detection of phishing websites. *IEEE Access*, *8*, 142532-142542. https://doi.org/10.1109/ACCESS.2020.3013699

[28] Al-Sarem, M., Saeed, F., Al-Mekhlafi, Z. G., Mohammed, B. A., Al-Hadhrami, T., Alshammari, M. T., Alreshidi, A., & Alshammari, T. S. (2021). An optimized stacking ensemble model for Phishing Websites Detection. *Electronics*, *10*(11), 1285. https://doi.org/10.3390/electronics10111285

[29] Adeyemo, V. E., Balogun, A. O., Mojeed, H. A., Akande, N. O., & Adewole, K. S. (2020, December). Ensemble-based logistic model trees for website phishing detection. In *International Conference on Advances in Cyber Security* (pp. 627-641). Springer, Singapore. https://doi.org/10.1007/978-981-33-6835-4_41

[30] *Sklearn.Feature_selection.RFECV*. (n.d.). Scikit-Learn. Retrieved April 2, 2022, from https://scikit-learn.org/stable/modules/generated/sklearn.feature_selection.RFECV.html

[31] Balogun, A. O., Adewole, K. S., Raheem, M. O., Akande, O. N., Usman-Hamza, F. E., Mabayoje, M. A., Akintola, A. G., Asaju-Gbolagade, A. W., Jimoh, M. K., Jimoh, R. G., & Adeyemo, V. E. (2021). Improving the phishing website detection using empirical analysis of Function Tree and its variants. *Heliyon*, *7*(7), e07437. https://doi.org/10.1016/j.heliyon.2021.e07437

[32] Khan, S. A., Khan, W., & Hussain, A. (2020, October). Phishing attacks and websites classification using machine learning and multiple datasets (a comparative analysis). In *International Conference on Intelligent Computing* (pp. 301-313). Springer, Cham. https://doi.org/10.1007/978-3-030-60796-8_26

[33] Harinahalli Lokesh, G., & BoreGowda, G. (2021). Phishing website detection based on effective machine learning approach. *Journal of Cyber Security Technology*, *5*(1), 1–14. https://doi.org/10.1080/23742917.2020.1813396